\documentstyle[preprint,aps]{revtex}

\begin{document}
\title{Generalized Bogoliubov Transformation, Fermion Field and Casimir Effect in a
Box }
\author{H. Queiroz$^{a}$ $,\,$J. C. da Silva$^{a,b}$, F.C. Khanna$^{c,d}$, M. Revzen$%
^{e}$, A. E. Santana$^{a}$.}
\address{${}^a$ Instituto de F\'\i sica, Universidade Federal da Bahia,\\
Campus de Ondina, 40210-340, Salvador, Bahia, Brasil.\\
$^b$ Centro Federal de Educa\c c\~ao Tecnol\'ogica da Bahia\\
Rua Em\'{i}dio Santos, 40000-900, Salvador, Bahia, Brasil\\
$^{c}$Physics Department, Theoretical Physics Institute,\\
University of Alberta, Edmonton, Alberta T6G 2J1 Canada\\
$^{d}$TRIUMF, 4004, Westbrook mall, Vancouver, British \\
Columbia V6T 2A3, Canada\\
$^{e}$Depart. of Physics, Technion - Institute of Technology \\
Haifa,32000, Israel. }
\date{\today }
\maketitle

\begin{abstract}
In this work a generalization of the Bogoliubov transformation is developed
to describe a space compactified fermionic field. The method is the fermion
counterpart of the formalism introduced earlier for bosons (J. C. da Silva,
A. Matos Neto, F.C. Khanna and A.E. Santana, Phys. Rev. A 66 (2002) 052101),
and is based on the thermofield dynamics approach. \ We analyze the
energy-momentum tensor for the Casimir effect of a free massless fermion
system in a $N$-dimensional box in contact with a heat bath. As a particular
situation we calculate the Casimir energy and pressure for the field in a
3-dimensional box of sizes $L_{1},L_{2},L_{3}.$ One interesting result is
that the attractive or repulsive nature of the Casimir pressure can change
depending on the rate among $L_{1},L_{2},L_{3}.$ This effect is exemplified
in the case of $L_{1}\rightarrow \infty ,$ and \ $L_{3}=L,L_{2}=0.1L.$
\end{abstract}

\section{Introduction}

In a recent work \cite{jura1} a generalization of the Bogoliubov
transformation was introduced to treat confined boson field in space
coordinates at finite temperature. The formalism is based on the thermofield
dynamics (TFD) approach \cite{ume1,ume2,ume4,oji1,kha5,gade1,gade2} and has
been applied to derive different aspects of the Casimir effect for the
electromagnetic field confined between two plates. In this context one
interesting physical result is brought about: the Casimir effect is
described by a process of condensation, thus shedding a new ingredient on
the nature \ of the quantum vacuum. In this paper we extend the approach
introduced in Ref. \cite{jura1} \ to fermion fields, and so we apply it to
the Casimir effect of a free massless fermionic field in a 3-dimensional box
at finite temperature.

The Casimir effect was first proposed by taking into account the effect of
the vacuum fluctuation of the electromagnetic field confined within two
plates with separation $L$, using the Dirichlet boundary conditions. The
result was an attractive force between the plates given by the negative
pressure $P=-\pi ^{2}/240L^{4}$ (we use natural units: $\hbar =c=1$) \cite
{casi1}. Over the decades the effect has been applied to different
geometries and physical conditions, enjoying a remarkable popularity \cite
{milon,mostep,mostep3,levin,seife,boyer1,milton1,plun,bordag,car1,car2,far1,lam1,roy,rev11,tesu1}
and raising interest, in particular, in the context of microelectronics \cite
{tec1,tec2}.

The effect of temperature was first studied by Lifshitz\cite{lif,pit} who
presented an alternative derivation for the Casimir force, including the
analysis of the dielectric nature of the material between the plates.
Actually, the effect of temperature on the interaction between the
conducting parallel plates may be significant for separations greater than $%
3\mu m$ \cite{mehra,mostep2,mann1,mann11,mann2,mann3}. For this physical
situation of plates, the full analysis of the thermal energy-momentum tensor
of the electromagnetic\ field was carried out by Brown and Maclay \cite
{brown}, performing the calculation of the Casimir free-energy by using the
Green's function (the local formulation) written in a conformally invariant
way \cite{plun,robaschik,takagi}. One of our proposal here is to derive the
fermionic counterpart of the Brown and Maclay's formula, but for a more
general situation of confinement: we consider not only the two plates, but
also the case of confinement within an $N$-dimensional box.

Casimir effect for a massless fermionic field is of great interest in
considering the structure of proton in particle physics; thus its physical
appeal. In particular, in the phenomenological MIT bag model \cite{bag},
quarks are confined in a small space region in such a way that there is no
fermionic current outside that region. The fermion field then fulfills the
so-called bag model boundary condition. The Casimir effect in such a small
region, of order $1.0fm,$ is important to define the process of
deconfinement in a heavy ion collision at Relativistic Heavy Ion Collider
(RHIC), giving rise to the quark-gluon plasma \cite{saito1}. The gluon field
contribution for the Casimir effect is, up to the color quantum numbers, the
same as for the electromagnetic field. For the quark field, the problem has
been often addressed only by considering the case of two parallel plates 
\cite{ravnd1,ravnd2,ravnd3,ravnd4,svai3,eli1}. Actually, as first
demonstrated by Johnson \cite{johns1}, for plates, the fermionic Casimir
force is attractive as in the case of the electromagnetic field. On the
other hand, depending on the geometry of the confinement, the nature of the
Casimir force can change. This is the case, for instance, of a sphere and
the Casimir-Boyer model, using mixed boundary conditions for the
electromagnetic field, such that the force is repulsive\cite
{ago1,ago11,ago2,ago3,ago4,jura2}. Therefore, the analysis considering
fermions in an Euclidian wave-guide (confinement in two-dimensions) and in a
3-dimensional box (confinement in 3-dimensions) may be of interest. We avoid
here the approach based on \ the sum of quantum modes, that was so important
in the calculation of the Brown and Maclay. Using, alternately, the method
developed in Ref. \cite{jura1}, we perform the calculation of the
non-trivial problem of the quark field in a finite volume with the
compactification in an $N$-dimensional box, at finite temperature.

In order to proceed with, we have to adapt the methodology of calculations
discussed earlier \cite{jura1}, to encompass a generalized Bogoliubov
transformation for fermion fields with the MIT bag model boundary condition;
equivalent to an antiperiodic boundary condition \cite{ravnd2}. This is
presented in Section 2, where we derive the energy-momentum tensor for the
fermion field at $T\neq 0$, resulting, as an example, in the
Stefan-Boltzmann law. This is the usual calculation of TFD for fermions;
however it can also be interpreted as a confinement in the time axis, in
such a way that the field is under anti-periodic boundary conditions. Using
then this methodology for a Euclidian geometry, we can envisage a space
compactification. (In fact, this possibility has also been explored in the
context of the Matsubara formalism \cite{jor11,jor12}.) The usual expression
of the Casimir effect is thus calculated, considering a proper modified
Bogoliubov transformation which will describe the confinement in the z-axis.
The familiarity with this kind of calculation for these already known
results will suggest to us a general form for a Bogoliubov transformation to
describe space compactification in arbitrary dimensions; a subject developed
in Section 3. In Section 4, we consider applications of the tensor derived
in Section 3. Then we present the main result of this paper: the conformally
invariant expression of the thermal energy-momentum tensor describing the
Casimir effect of a massless fermionic field confined in a 3-dimensional
box. Some particular situations are analyzed, resulting for instance that
the attractive or repulsive nature of the Casimir force can change in
accordance with the rate among the sizes of the box. This situation is
explicitly treated considering the case of wave guide obtained from a box of
sizes $L_{1},L_{2},L_{3},$ such that $L_{1}\rightarrow \infty ,$ and \ $%
L_{3}=L,L_{2}=0.1L.$ In Section 5 our concluding remarks are presented.

\section{Bogoliubov transformation and compactification}

The general approach to be used here can be addressed through the following
prescription, taken as a generalization of the TFD formalism \cite
{jura1,ume1,ume2} . Given an arbitrary set of operators, say ${\cal V}$,
with elements denoted by $A_{i},i=1,...,n$, there exists a mapping \
describing a doubling in the degrees of freedom defined by $\tau :{\cal V}%
\rightarrow {\cal V}$, denoted by $\tau A\tau ^{-1}=\widetilde{A},$
satisfying the following conditions 
\begin{eqnarray}
(A_{i}A_{j})\widetilde{} &=&\widetilde{A}_{i}\widetilde{A}_{j},
\label{c1til2} \\
(cA_{i}+A_{j})\widetilde{} &=&c^{\ast }\widetilde{A}_{i}+\widetilde{A}_{j},
\label{c1til3} \\
(A_{i}^{\dagger })\widetilde{} &=&(\widetilde{A}_{i})^{\dagger },
\label{c1til4} \\
(\widetilde{A}_{i})\widetilde{} &=&A_{i}.  \label{c1til5}
\end{eqnarray}
These properties are called {\em the tilde (or dual) conjugation rules} in
TFD. The doubling in the Hilbert space has a new vacuum denoted by $|0,%
\widetilde{0}\rangle $. Consider $\alpha =(\alpha _{0},\alpha _{1},\alpha
_{2},\alpha _{3},...)$ a set of $c$-numbers associated with macroscopic
parameters of the system as such: temperature ($\beta =1/T),$ and three
possible parameters describing the spatial confinement (for instance the
sizes $L_{1}$, $L_{2}$ and $L_{3}$ of a box in space). Then there exists a
Bogoliubov transformation given by 
\begin{equation}
B(\alpha )=\left( 
\begin{array}{cc}
\,\,\,\,\,u(\alpha ) & -v(\alpha ) \\ 
v(\alpha ) & \,\,u(\alpha )
\end{array}
\right) ,  \label{jurah3}
\end{equation}
with $\,u^{2}(\alpha )+v^{2}(\alpha )=1$.

For an arbitrary operator in ${\cal V}$, we use the doublet notation \cite
{ume1}

\begin{eqnarray}
(A^{a}) &=&\left( 
\begin{array}{c}
A(\alpha ) \\ 
\widetilde{A}^{\dagger }(\alpha )
\end{array}
\right) =B(\alpha )\left( 
\begin{array}{c}
A \\ 
\widetilde{A}^{\dagger }
\end{array}
\right) ,\,\,\,  \label{cs1} \\
(\,A^{a})^{\dagger } &=&\left( A^{\dagger }(\alpha )\,\,,\,\,\widetilde{A}%
(\alpha )\right) .\,\,  \label{cs9}
\end{eqnarray}
This kind of notation is useful to calculate the propagator for the confined
field.

Here we are concerned with the energy-momentum tensor for a massless
fermionic field given by \cite{ravnd1,fermion} 
\begin{eqnarray}
T^{\mu \nu }(x) &=&\langle 0|i\overline{\psi }(x^{\prime })\gamma ^{\mu
}\partial ^{\nu }\psi (x)|0\rangle |_{x^{\prime }\rightarrow x}
\label{adee1} \\
&=&\gamma ^{\mu }\partial ^{\nu }S(x-x^{\prime })|_{x^{\prime }\rightarrow x}
\label{adee2} \\
&=&-4i\partial ^{\mu }\partial ^{\nu }G_{0}(x-x^{\prime })|_{x^{\prime
}\rightarrow x},  \label{adee3}
\end{eqnarray}
where $S(x-x^{\prime })=-i\langle 0|T[\psi (x)\overline{\psi }(x^{\prime
})]|0\rangle $ and 
\begin{eqnarray*}
G_{0}(x) &=&\frac{-1}{(2\pi )^{4}}\int d^{4}k\,\,e^{-ik\cdot x}G_{0}(k) \\
&=&\frac{-i}{(2\pi )^{2}}\frac{1}{x^{2}-i\varepsilon },
\end{eqnarray*}
such that 
\[
G_{0}(k)=\frac{1}{k^{2}+i\varepsilon }, 
\]
and the Minkowski metric with a signature $(+---)$. With $T^{\mu \nu }(x)$,
we can introduce the confined ($\alpha $-)energy-momentum tensor ${\cal T}%
^{\mu \nu (ab)}(x;\alpha )$ defined by 
\begin{equation}
{\cal T}^{\mu \nu (ab)}(x;\alpha )=\langle T^{\mu \nu (ab)}(x;\alpha
)\rangle -\langle T^{\mu \nu (ab)}(x)\rangle ,  \label{ade122}
\end{equation}
where $T^{\mu \nu (ab)}(x;\alpha )$ is a function of the field operators $%
\psi (x;\alpha )$,$\ \widetilde{\psi }(x;\alpha )$ and $\langle \cdot \cdot
\cdot \rangle =\langle 0,\widetilde{0}|\cdot \cdot \cdot |0,\widetilde{0}%
\rangle .$ The physical $\alpha $-tensor is given by the component ${\cal T}%
^{\mu \nu (11)}(x;\alpha ).$ Let us then work out this tensor.

Considering the TFD prescription \cite{ume1,sout1}, we have 
\[
S^{(ab)}(x-x^{\prime })=\left( 
\begin{array}{cc}
S(x-x^{\prime }) & 0 \\ 
0 & \widetilde{S}(x-x^{\prime })
\end{array}
\right) , 
\]
with $\widetilde{S}(x-x^{\prime })=-S^{\ast }(x^{\prime }-x)$. As a result,
from Eq.(\ref{ade122}), \ we have 
\begin{equation}
{\cal T}^{\mu \nu (ab)}(x;\alpha )=-4i\partial ^{\mu }\partial ^{\nu
}[G_{0}^{(ab)}(x-x^{\prime };\alpha )-G_{0}^{(ab)}(x-x^{\prime
})]_{x^{\prime }\rightarrow x},  \label{jura2}
\end{equation}
corresponding to a change in Eq.(\ref{adee2}), $S^{(ab)}(x-x^{\prime })$ by $%
S(x-x^{\prime })$. The Green's functions in Eq.(\ref{jura2}) are given by 
\[
G_{0}^{(ab)}(x-x^{\prime })=\frac{-1}{(2\pi )^{4}}\int
d^{4}k\,\,G_{0}^{(ab)}(k)\,\,e^{-ik\cdot (x-x^{\prime })}, 
\]
where 
\[
G_{0}^{(ab)}(k)=\left( 
\begin{array}{cc}
G_{0}(k) & 0 \\ 
0 & G_{0}^{\ast }(k)
\end{array}
\right) , 
\]
and the $\alpha $-counterpart is 
\begin{equation}
G_{0}^{(ab)}(x-x^{\prime };\alpha )=\frac{-1}{(2\pi )^{4}}\int
d^{4}k\,\,G_{0}^{(ab)}(k;\alpha )\,\,e^{-ik\cdot (x-x^{\prime })},
\label{jurah5}
\end{equation}
with 
\[
G_{0}^{(ab)}(k;\alpha )=B_{k}^{-1(ac)}(\alpha
)G_{0}^{(cd)}(k)B_{k}^{(db)}(\alpha ), 
\]
where $B_{k}^{(ab)}(\alpha )$ is the Bogoliubov transformations given in Eq.(%
\ref{jurah3}). Explicitly, the components of $G_{0}^{(ab)}(k;\alpha )$ are
given by 
\begin{eqnarray*}
\,\,G^{11}(k;\alpha ) &=&G_{0}(k)+v_{k}^{2}(\alpha )[G_{0}^{\ast
}(k)-G_{0}(k)], \\
\,G^{12}(k;\alpha ) &=&G^{21}(k;\alpha )=v_{k}(\alpha )[1-v_{k}^{2}(\alpha
)]^{1/2}[G_{0}^{\ast }(k)-G_{0}(k)], \\
G^{22}(k;\alpha ) &=&G_{0}^{\ast }(k)+v_{k}^{2}(\alpha
)[G_{0}(k)-G_{0}^{\ast }(k)].
\end{eqnarray*}
The physical quantities are derived from the component $G^{11}(k;\alpha )$.

Let us consider a simple situation in which $\alpha \equiv \beta =1/T$ . In
this case $v_{k}(\beta )$ is defined through the fermion number
distribution, that is 
\[
v_{k}(\beta )=\frac{e^{-\beta k_{0}/2}}{[1+e^{-\beta k_{0}}]^{1/2}}. 
\]
Observe that we can write 
\begin{equation}
v_{k}^{2}(\beta )=\sum_{l=1}^{\infty }(-1)^{l+1}e^{-\beta k_{0}l};
\label{jurah8}
\end{equation}
leading to the thermal Green's function, 
\[
G_{0}^{11}(k;\beta )=G_{0}(k)+\sum_{l=1}^{\infty }(-1)^{l+1}e^{-\beta
k_{0}l}[G_{0}^{\ast }(k)-G_{0}(k)]. 
\]
Using this result in Eq.(\ref{jurah5}) we derive 
\[
G_{0}^{11}(x-x^{\prime };\beta )=G_{0}(x-x^{\prime })+\sum_{l=1}^{\infty
}(-1)^{l+1}[G_{0}^{\ast }(x^{\prime }-x-i\beta l\widehat{n}%
_{0})-G_{0}(x-x^{\prime }-i\beta l\widehat{n}_{0})], 
\]
where $\widehat{n}_{0}=(1,0,0,0)$ is a time-like vector. Therefore, from Eq.(%
\ref{jura2}), we find 
\[
{\cal T}^{\mu \nu (11)}(\beta )=4i\sum_{l=1}^{\infty }(-1)^{l+1}\partial
^{\mu }\partial ^{\nu }[G_{0}(x^{\prime }-x+i\beta l\widehat{n}%
_{0})+G_{0}(x-x^{\prime }-i\beta l\widehat{n}_{0})]|_{x^{\prime }\rightarrow
x}. 
\]
Performing the covariant derivatives, this expression reads 
\begin{equation}
{\cal T}^{\mu \nu (11)}(\beta )=\frac{8}{(2\pi )^{2}}\sum_{l=1}^{\infty
}(-1)^{l}\left[ \frac{2g^{\mu \nu }-8\widehat{n}_{0}^{\mu }\widehat{n}%
_{0}^{v}}{(\beta l)^{4}}\right] .  \label{jurah9}
\end{equation}

Well known results for thermal fermionic fields can be derived from this
tensor. For instance, the internal energy is given by $E(T)={\cal T}%
^{00(11)}(\beta ),$ that is, 
\[
E(T)=\frac{7}{4}\frac{\pi ^{2}}{15}T^{4}, 
\]
where we have used the Riemann zeta-function \cite{ravnd1} 
\[
\varsigma (4)=\sum_{l=1}^{\infty }(-1)^{l}\frac{1}{l^{4}}=-\frac{7}{8}\frac{%
\pi ^{4}}{90}. 
\]

As another application, we derive the Casimir effect, by following the above
calculations. In this case, instead of Eq.(\ref{jurah8}), we write $\alpha
=i\alpha _{3}=ia$ 
\[
v_{k}^{2}(a)=\sum_{l=1}^{\infty }(-1)^{l+1}e^{-ik_{3}al}, 
\]
and use $\widehat{n}_{3}=(0,0,0,1)$ a space-like vector. As a consequence we
derive 
\begin{equation}
{\cal T}^{\mu \nu (11)}(a)=\frac{8}{(2\pi )^{2}}\sum_{l=1}^{\infty }(-1)^{l}%
\left[ \frac{2g^{\mu \nu }+8\widehat{n}_{3}^{\mu }\widehat{n}_{3}^{\nu }}{%
(al)^{4}}\right] .  \label{jurah10}
\end{equation}
Resulting in a Casimir energy and pressure given, respectively, by 
\[
E_{c}(a)={\cal T}^{00(11)}(a)=-\frac{7}{4}\frac{\pi ^{2}}{45a^{4}}, 
\]
\begin{equation}
P_{c}(a)={\cal T}^{33(11)}(a)=-3\frac{7}{4}\frac{\pi ^{2}}{45a^{4}},
\label{jurah202}
\end{equation}
here$\ a=2L$, where $L$ is the separation of the plates. This is needed (as
in the case of bosons) in order to fix the antiperiodic boundary conditions
on the propagator, $S(x-x^{\prime })$.

\section{Compactification in Higher Dimensions}

In this section we calculate the Casimir effect for the massless fermions
within an $N$-dimensional (space) box at finite temperature. We proceed then
by generalizing the results for the temperature and the Casimir effect at
zero temperature which were derived in the last section. Supported by those
calculations, we consider a generalization of $v(\alpha )$ as given in Eq.(%
\ref{jurah8}) by writing $\alpha \rightarrow \alpha =(\alpha _{0},\alpha
_{1},\alpha _{2},...,\alpha _{N})$, $\widehat{n}_{0}=(1,0,0,0,...)$, $%
\widehat{n}_{1}=(0,1,0,0,...)$, $\widehat{n}_{2}=(0,0,1,0,...),...,$ $%
\widehat{n}_{N}=(0,0,0,...,1)$, vectors in the ($N+1)$-dimensional Minkowski
space, such that 
\[
v_{k}^{2}(\alpha )=\sum_{l_{0},l_{1},...,l_{N}=0^{\prime }}^{\infty
}(-1)^{l_{0}+l_{1}+...+l_{N}+N_{c}}\,\exp \{i%
\mathop{\textstyle\sum}%
_{i=0}^{N}\alpha _{i}l_{i}k_{i}\},
\]
where $N_{c}$ is the number of nonzero component of $\alpha _{\mu }$ in set
of sum (for instance for each sum of the type $\sum_{l_{j}=1}$ , then $%
N_{c}=1$, for sums of type $\sum_{l_{k}l_{j}=1},$ then $N_{c}=2,$ and so
on), and the symbol $0^{\prime }$ in the sum means that the situation in
with $l_{0}=l_{1}=...=l_{N}=0$ is excluded. Using this $v_{k}^{2}(\alpha )$
and the procedure delineated in Section 2, we obtain 
\begin{eqnarray*}
{\cal T}^{\mu \nu (11)}(\alpha ) &=&4i\sum_{l_{0},l_{1},...,l_{N}=0^{\prime
}}^{\infty }(-1)^{l_{0}+l_{1}+...+l_{N}+N_{c}}\,\,\partial ^{\mu }\partial
^{\nu } \\
&&\times \lbrack G_{0}(x^{\prime }-x+\alpha _{0}l_{0}\widehat{n}%
_{0}-\sum_{i=1}^{N}a_{i}l_{i}\widehat{n}_{i})+G_{0}(x-x^{\prime }-\alpha
_{0}l_{0}\widehat{n}_{0}-\sum_{i=1}^{N}a_{i}l_{i}\widehat{n}%
_{i})]|_{x^{\prime }\rightarrow x},
\end{eqnarray*}
resulting in 
\begin{eqnarray}
{\cal T}^{\mu \nu (11)}(\alpha ) &=&\frac{-8}{(2\pi )^{2}}%
\sum_{l_{0},l_{1},...,l_{N}=0^{\prime }}^{\infty
}(-1)^{l_{0}+l_{1}+...+l_{N}+N_{c}}\left\{ \frac{1}{[\sum_{i=1}^{N}(\alpha
_{i}l_{i})^{2}-(\alpha _{0}l_{0})^{2}]^{2}}\right.   \nonumber \\
&&\left. \times \left[ 2g^{\mu \nu }+8\frac{\sum_{i,j=1}^{N}(\alpha
_{i}l_{i})(\alpha _{j}l_{j})\widehat{n}_{i}^{\mu }\widehat{n}_{j}^{\nu
}+(\alpha _{0}l_{0})^{2}\widehat{n}_{0}^{\mu }\widehat{n}_{0}^{\nu }}{%
\sum_{i=1}^{N}(\alpha _{i}l_{i})^{2}-(\alpha _{0}l_{0})^{2}}\right] \right\}
.  \label{tan1}
\end{eqnarray}
Notice that the results given by Eqs.(\ref{jurah9}) and (\ref{jurah10}) are
particular cases of the energy-momentum tensor given by Eq.(\ref{tan1}).
Another important aspect is that ${\cal T}^{\mu \nu (11)}(\alpha )$ is
traceless, as it should be.

\section{Thermal Casimir Effect in a Box}

In the case of a $3$-dimensional closed box, considering the temperature
effect, we have $\alpha _{0}=i\beta ,$ $\alpha _{i}=2L_{i}(i=1,2,3)$ where $%
L_{i}$ stands for the size of the $i$-th direction of \ the box, and $%
N_{c}=4 $ . Using Eq.(\ref{tan1}) in a $(3+1)$ Minkowski space, we obtain 
\begin{eqnarray}
{\cal T}^{\mu \nu (11)}(\alpha ) &=&\frac{-8}{(2\pi )^{2}}%
\sum_{l_{0},l_{1},l_{2},l_{3}=0^{\prime }}^{\infty
}(-1)^{l_{0}+l_{1}+l_{2}+l_{3}+N_{c}}\left\{ \frac{1}{%
[\sum_{i=1}^{3}(2L_{i}l_{i})^{2}+(\beta l_{0})^{2}]^{2}}\right.  \nonumber \\
&&\times \left. \left[ 2g^{\mu \nu }+8\frac{%
\sum_{i,j=1}^{3}(2L_{i}l_{i})(2L_{j}l_{j})\widehat{n}_{i}^{\mu }\widehat{n}%
_{j}^{\nu }-(\beta l_{0})^{2}\widehat{n}_{0}^{\mu }\widehat{n}_{0}^{\nu }}{%
\sum_{i=1}^{3}(2L_{i}l_{i})^{2}+(\beta l_{0})^{2}}\right] \right\} .
\label{tan2}
\end{eqnarray}
For a cubic box with $L_{i}=L,$ for all $i=1,2,3$, the Casimir energy ($%
E_{cb}(L))$ and the Casimir pressure ($P_{cb}(L))$ along the direction $i=3$
at zero temperature are, respectively, given by 
\begin{eqnarray}
E_{cb}(L) &=&{\cal T}^{00(11)}(L)  \nonumber \\
&=&\frac{8}{(2\pi )^{2}}\sum_{l_{0},l_{1},l_{2},l_{3}=0^{\prime }}^{\infty
}(-1)^{l_{1}+l_{2}+l_{3}+N_{c}}\left\{ \frac{2}{%
[\sum_{i=1}^{3}(2Ll_{i})^{2}]^{2}}\right\}  \label{hebe1}
\end{eqnarray}
and 
\begin{eqnarray}
P_{cb}(L) &=&{\cal T}^{33(11)}(L)  \nonumber \\
&=&\frac{-8}{(2\pi )^{2}}\sum_{l_{1},l_{2},l_{3}=0^{\prime }}^{\infty
}(-1)^{l_{1}+l_{2}+l_{3}+N_{c}}\left\{ \frac{%
2\sum_{i=1}^{3}(2Ll_{i})^{2}-8(2Ll_{3})^{2}}{%
[\sum_{i=1}^{3}(2Ll_{i})^{2}]^{3}}\right\} .  \label{hebe2}
\end{eqnarray}

As another possibility for compactification, let calculate the expression
for the Casimir pressure at zero temperature, considering the confinement in
two dimensions (four plates), that is the Casimir effect for a wave guide,
with $L_{1}\rightarrow \infty $. In this case, using Eq.(\ref{tan1}), we
have the Casimir pressure, $P_{cg}(L),$ along the direction z (we consider
the confinement in y- and z- axes, with $L_{2}=L_{3}=L$) 
\begin{eqnarray}
P_{cg}(L) &=&\frac{8}{(2\pi )^{2}}\sum_{l_{2},l_{3}=0^{\prime }}^{\infty
}(-1)^{l_{2}+l_{3}+N_{c}}\left\{ 2\frac{(2L_{2}l_{2})^{2}-3(2L_{3}l_{3})^{2}%
}{[(2L_{2}l_{2})^{2}+(2L_{3}l_{3})^{2}]^{3}}\right\}   \nonumber \\
&=&\frac{8}{(2\pi )^{2}}\sum_{l_{2}=1}^{\infty }\sum_{l_{3}=1}^{\infty
}\left\{ (-1)^{l_{2}}(-1)^{l_{3}}\frac{2((2L_{2}\left( l_{2}\right)
)^{2}-6(2L_{3}\left( l_{3}\right) )^{2}}{((2L_{2}\left( l_{2}\right)
)^{2}+(2L_{3}\left( l_{3}\right) )^{2})^{3}}\right\}   \nonumber \\
&&-\frac{8}{(2\pi )^{2}}\sum_{l_{2}=1}^{\infty }\left\{ (-1)^{l_{2}}\frac{%
2((2L_{2}\left( l_{2}\right) )^{2}}{((2L_{2}\left( l_{2}\right) )^{2})^{3}}%
\right\} +\frac{8}{(2\pi )^{2}}\sum_{l_{3}=1}^{\infty }\left\{ (-1)^{l_{3}}%
\frac{6(2L_{3}\left( l_{3}\right) )^{2}}{((2L_{3}\left( l_{3}\right)
)^{2})^{3}}\right\} .  \label{hebe33}
\end{eqnarray}
The consistency of this formula can be verified by taking limits. For
instance, from the Casimir pressure for the box, Eq.(\ref{hebe2}), we can
recover Eq.(\ref{hebe2}) or Eq.(\ref{jurah202}) by taking, respectively, the
limits $L_{1}\rightarrow \infty $ and $L_{1},L_{2}\rightarrow \infty .$

In Figure 1 the full line is the plot of $P_{c}(L),$ the Casimir pressure
due to two parallel plates separated by a distance $L$ as given in Eq.(\ref
{jurah202}); the dashed line is the Casimir pressure $P_{cb}(L)$ for a cubic
box as given in Eq.(\ref{hebe2}) with $L_{1}=L_{2}=L_{3}=L$ (the plot of
Casimir pressure, $P_{cg}(L),$ for a wave guide obtained from Eq.(\ref
{hebe33}) with $L_{1}\rightarrow \infty $ $,L_{2}=L_{3}=L,$ is virtually the
same as $P_{c}(L)$ in the scale of \ Figure 1). In Figure 2 there is the
plot of $P_{cg}(L)$ derived from Eq.(\ref{hebe2}) with $L_{1}\rightarrow
\infty ,$ and \ $L_{3}=L,L_{2}=0.1L.$\ Observe that the Casimir pressure is
always positive in this case.

\section{Concluding Remarks}

In this paper we have presented a generalization of the Bogoliubov
transformation in order to describe a massless fermion field compactified in
an $N$-dimensional box at finite temperature. We write the (traceless)
energy-momentum tensor from which we calculate and compare explicit
expressions for the Casimir pressure, corresponding to different cases of
confinement. The Casimir pressure for the case of two parallel plates and
the cubic box are negative, imposing then an attractive force on the plates
along the direction of the analysis. However, in the case for different
rates among the sizes of the box, as in the example of the wave guide (the
confinement in two directions) treated in Figure 2, the pressure is
positive, representing a repulsive force among the plates. This repulsive
force can have a direct influence in the description of quark deconfinement,
pointing to an adverse effect when we compare it with the usual calculation
of the Casimir force using the two plates (confinement along one dimension
only), in which the pressure is attractive.

Another aspect, worthy of noting, is the simplicity of calculations, that
can be observed from the known results for the fermion Casimir effect; see
for instance \cite{milton1,saito1}. This is so since we have avoided usual
procedures, as the intricate method based on the sum of the quantum modes of
the fields, satisfying some given boundary conditions. Indeed, instead of
the sum of modes, we have used the Bogoliubov transformation to define the
Casimir effect as a kind of condensation procedure of the fermion field (in
a similar fashion as was carried out for the case of bosons\cite{jura1}).
Taking advantages of these practical proposals, the method developed here
can be useful for calculations involving other geometries such as spherical
or cylindrical symmetries. This analysis will be developed in more detail
elsewhere.

\begin{description}
\item  {\bf Acknowledgments: } This work was supported by CNPQ of Brazil, by
NSERC of Canada, by the Technion-Haifa University Joint Research Fund, and
by the Fund for Promotion of Sponsored Research at the Technion, Israel.
\end{description}

\newpage

Figure 1

Full line: Casimir pressure, $P_{c}$, for two parallel plates separeted by a
distance $L;$dashed line: Casimir pressure, $P_{cb}$, for a box with sizes
specifyed by $L_{1}=L_{2}=L_{3}=L$ 
\begin{eqnarray*}
&& \\
&&
\end{eqnarray*}

Figure 2

\bigskip

Casimir pressure, $P_{cg}$, for a wave guide obtained from a box with $%
L_{1}\rightarrow \infty ,$ and \ $L_{3}=L,L_{2}=0.1L$


\begin{references}
\bibitem{jura1}  J. C. da Silva, F.C. Khanna, A. Matos Neto  and A. E.
Santana, Phys. Rev. A {\bf 66} (2002) 052101.

\bibitem{ume1}  H. Umezawa, {\em Advanced Field Theory: Micro, Macro and
Thermal Physics}{\it \ }(AIP, New York, 1993).

\bibitem{ume2}  Y. Takahashi and H. Umezawa, Coll. Phenomena {\bf 2} (1975)
55 (Reprinted in Int. J. Mod. Phys. 10 (1996) 1755).

\bibitem{ume4}  H. Umezawa, H. Matsumoto and M. Tachiki,{\em \ Thermofield
Dynamics and Condensed States} (North-Holland, Amsterdan, 1982).

\bibitem{oji1}  I. Ojima, Ann. Phys. (N. Y.) {\bf 137 }(1981) 1.

\bibitem{kha5}  A. E. Santana, A. Matos Neto, J. D. M. Vianna and F. C.
Khanna, Physica A {\bf 280} (2000) 405.

\bibitem{gade1}  M. C. B. Abdalla, A. L. Gadelha and I. V. Vancea, Phys. Rev.
D {\bf 64} (2001) 086005.

\bibitem{gade2}  M. C. B. Abdalla, A. L. Gadelha and I. V. Vancea, Int. J.
Mod. Phys. A {\bf 18} (2003) 2109.

\bibitem{casi1}  H. B. G. Casimir, Proc. Ned. Akad. Wet. B {\bf 51} (1948)
793.

\bibitem{milon}  P. W. Milonni, {\em The Quantum Vacuum} (Academic, Boston,
1993).

\bibitem{mostep}  V. M. Mostepanenko and N.N. Trunov, {\em The Casimir
Effect and its Applications} (Clarendon, Oxford, 1997).

\bibitem{mostep3}  M. Bordag, U. Mohideed and V. M. Mostepanenko, {\em New
Developments in Casimir Effect}, quant-ph/0106045, Phys. Rep. {\bf 353}
(2001) 1.

\bibitem{levin}  F. S. Levin and D.A. Micha (Eds.), {\em Long Range Casimir
Forces} (Plenum, New York, 1993).

\bibitem{seife}  C. Seife, Science {\bf 275} (1997) 158.

\bibitem{boyer1}  T. H. Boyer, Am. J. Phys. {\bf 71} (2003) 990.

\bibitem{milton1}  K. A. Milton, {\em The Casimir Effect: Physical
Manifestations of Zero Point Energy}, hep-th/9901011.

\bibitem{plun}  G. Plunien, B. M\"{u}eller and W. Greiner, Phys. Rep. {\bf %
134 }(1986) 87.

\bibitem{bordag}  M. Bordag (Ed.), {\em The Casimir Effect 50 Years Later}
(World Scientific, Singapore, 1999).

\bibitem{car1}  F. Caruso, N. P. Neto, B. F. Svaiter and N. F. Svaiter,
Phys. Rev. D {\bf 43 }(1991) 1300.

\bibitem{car2}  F. Caruso, R. de Paola and N. F. Svaiter, Int. J. Mod. Phys.
A {\bf 14 }(1999) 2077.

\bibitem{far1}  M. V. Cougo-Pinto, C. Farina and A. Ten\'{o}rio, Braz. J.
Phys. {\bf 29 }(1999) 371.
\bibitem{lam1}  S. K. Lamoreaux, Am. J. Phys.{\bf \ 67} (1999) 850.

\bibitem{roy}  U. Mohideen and A. Roy, Phys. Rev. Lett. {\bf 81} (1998) 4549.

\bibitem{rev11}  O. Kenneth, I. Klich, A. Mann and M. Revzen, Phys. Rev.
Lett. {\bf 89} (2002) 033001.

\bibitem{tesu1}  T. Maruyama, K. Tsushima and A. Faessler, Nucl. Phys. A 
{\bf 537 }(1992) 303.

\bibitem{tec1}  F. Serry, D. Walliser and G. J. Maclay, J. Appl. Phys. {\bf %
84 }(1998) 2501.

\bibitem{tec2}  E. Buks and M. L. Roukes, Phys. Rev. B {\bf 63 }(2001)
033402.

\bibitem{lif}  E. M. Lifshitz, Sov. Phys. JETP {\bf \ 2} (1956) 73.

\bibitem{pit}  I. E. Dzyaloshinskii, E. M. Lifshitz and L. P. Pitaevskii,
Adv. Phys.{\bf 10 }(1961) 165.

\bibitem{mehra}  J. Mehra, Physica, {\bf 37} (1967) 145.

\bibitem{mostep2}  G. L. Klimchitskaya and V. M. Mostepanenko, Phys. Rev. A 
{\bf 63 }(2001) 062108.

\bibitem{mann1}  M. Revzen, R. Opher, M. Opher and A. Mann, Europhys. Lett. 
{\bf 38} (1997) 245 (1997).

\bibitem{mann11}  M. Revzen, R. Opher, M. Opher and A. Mann, J. Phys. A:
Math and Gen. {\bf 30} (1997) 7783.

\bibitem{mann2}  M. Revzen and A. Mann, {\em Casimir Effect - The Classical
Limit}, quant-ph/9803059.

\bibitem{mann3}  J. Feinberg, A. Mann and M. Revzen, Ann. Phys.(NY) 288
(2001) 103.

\bibitem{brown}  L. S. Brown and G. J. Maclay, Phys. Rev. {\bf 184} (1969)
1272.

\bibitem{robaschik}  D. Robaschik, K. Scharnhorst and E. Wieczorek, Ann.
Phys.(N.Y.) {\bf 174} (1987) 401.

\bibitem{takagi}  S. Tadaki and S. Takagi, Prog. Theor. Phys. {\bf 75 }%
(1982) 262.

\bibitem{bag}  A. Chodos, R. L. Jaffe, K. Johnson, C. B. Thorn and V. F.
Weisskopf, Phys. Rev. D {\bf 9} (1974) 3471.

\bibitem{saito1}  K. Saito, Z. Phys. C {\bf 50} (1991) 69.

\bibitem{ravnd1}  S. A. Gundersen and F. Ravndal, Ann. Phys. (N.Y.) {\bf 182}
(1988) 90.

\bibitem{ravnd2}  C. A. L\"{u}tken and F. Ravndal, J. Phys. A: Math. Gen. 
{\bf 21} (1988) L793.

\bibitem{ravnd3}  C. A. L\"{u}tken and F. Ravndal, J. Phys. G: Nucl. Phys. 
{\bf 10 }(1984) 123.

\bibitem{ravnd4}  F. Ravndal and D. Tollefsen, Phys. Rev. D {\bf 40} (1989)
4191.

\bibitem{svai3}  R. D. M. De Paola, R. B. Rodrigues and N. F. Svaiter, Mod.
Phys. Lett. A {\bf 14} (1999) 2353.

\bibitem{eli1}  E. Elizalde, F. C. Santos and A. C. Tort, Int. J. Mod. Phys.
A {\bf 18} (2003) 1761, hep-th/0206114.

\bibitem{johns1}  K. Johnson, Acta Phys. Pol. B {\bf 6} (1975) 865.

\bibitem{ago1}  L. A. Ferreira, A. H. Zimerman and J. R. Ruggiero, {\em %
Casimir Effect for Closed Cavities with Conducting and Permeable Walls, }%
Instituto de F\'{i}sica Te\'{o}rica, UNESP, Preprint IFT-P-15/80.

\bibitem{ago11}  W. Lukosz, Physica {\bf 56} (1971) 109

\bibitem{ago2}  J. Ambj$\phi $rn and S. Wolfram, Ann. Phys. (N.Y.) {\bf 147}
(1983) 1.

\bibitem{ago3}  J. Ambj$\phi $rn and S. Wolfram, Ann. Phys. (N.Y.) {\bf 147}
(1983) 33.

\bibitem{ago4}  T. H. Boyer, Phys. Rev. A {\bf 9} (1974) 2078.

\bibitem{jura2}  J. C. da Silva, A. Matos Neto, H. Queiroz Pl\'{a}cido, M.
Revzen and A. E. Santana, Physica A {\bf 292} (2001) 411.

\bibitem{jor11}  A. P. C. Malbouisson and J. M. C. Malbouisson, J. Phys. A:
Math. Gen.{\bf \ 35} (2001) 2263.

\bibitem{jor12}  A. P. C. Malbouisson, J. M. C. Malbouisson and A. E.
Santana, Nucl. Phys. B {\bf 631 }(2002) 83.

\bibitem{fermion}  C. Itzykson and J. B. Zuber, {\em Quantum Field Theory (}%
McGrow-Hill, New York, 1980).

\bibitem{sout1}  K. Soutome, Z. Phys. C {\bf 40} (1988) 479.
\end{references}
\end{document}